\documentclass[twocolumn,showpacs,floats,prl,aps]{revtex4}

\usepackage[dvips]{graphicx}
\usepackage{amsfonts}
\usepackage{amsmath}
\usepackage{amssymb}
\usepackage{exscale}
\usepackage{eufrak}
\usepackage{afterpage}		

\begin{document}



\title{Anisotropic dressing of charge-carriers in the electron-doped
        cuprate superconductor Sm$_{1.85}$Ce$_{0.15}$CuO$_{4}$ from
        angle-resolved photoemission measurements}

\author{A. F. Santander-Syro,$^{1,2}$ 
		T. Kondo,$^{3}$
		J. Chang,$^{4}$
		A. Kaminski,$^{3}$
		S. Pailh\`es,$^{4,5}$ 
		M. Shi,$^{6}$
		L. Patthey,$^{6}$
		A. Zimmers,$^{7}$
		B. Liang,$^{7}$
		P. Li,$^{7}$ and
		R. L. Greene$^{7}$
		}
		
\address{$^{1}$Laboratoire Photons Et Mati\`ere, UPR-5 CNRS, ESPCI, 
		10 rue Vauquelin, 75231 Paris cedex 5, France}
\address{$^{2}$Labratoire de Physique des Solides, UMR-8502 CNRS, Universit\'e Paris-Sud, 
		91405 Orsay, France}
\address{$^{3}$Ames Laboratory and Department of Physics and Astronomy, Iowa State University,
		Ames, IA 50011}
\address{$^{4}$Laboratory for Neutron Scattering, ETH Zurich and Paul Scherrer Institute, 
		CH-5232 Villigen PSI, Switzerland}		
\address{$^{5}$Laboratoire L\'eon Brillouin, CEA-CNRS, CEA-Saclay, 91191 Gif-sur-Yvette, France}
\address{$^{6}$Swiss Light Source, Paul-Scherrer Institut, CH-5232 Villigen, Switzerland}
\address{$^{7}$Center for Nanophysics and Advanced Materials, Department of Physics, 
		University of Maryland, College Park, MD 20742}
		 
\date{\today}


\begin{abstract}
Angle-resolved photoemission measurements on the 
electron-doped cuprate Sm$_{1.85}$Ce$_{0.15}$CuO$_{4}$  
evidence anisotropic dressing of charge-carriers due to many-body interactions.
Most significantly, the scattering rate along the zone boundary
saturates for binding energies larger than  $\sim 200$~meV,
while along the diagonal direction it increases 
nearly linearly with the binding energy in the energy range $\sim 150-500$~meV.
These results indicate that many-body interactions along the diagonal direction
are strong down to the bottom of the band, while along the zone-bounday
they become very weak at energies above $\sim 200$~meV.
\end{abstract}


\maketitle  


%
Strong interactions in many-body systems lead to a rich variety
of phenomena, whose understanding is a central question in modern
physics. In condensed matter, 
cuprates are a paradigmatic example --and a continuing challenge-- 
of the physics of strong electronic correlations. 
Understanding how strong interactions in cuprates affect
their electronic structure is important to explain  
their properties, including the question of the pairing mechanism.
In fact, the coupling of the carriers to elementary excitations 
affect the carriers' band dispersion 
and energy-dependent scattering rate ($1/\tau$), which can be
obtained respectively from the positions and widths of the spectral peaks
in angle-resolved photoemission spectroscopy (ARPES) experiments~\cite{Damascelli-Review}.

In hole-doped (h-doped) cuprates, 
$1/\tau$ has been mainly studied
in the vicinity of the zone diagonal (D),
where the superconducting (SC) gap vanishes. Here, the 
carriers' dispersion and scattering rate display a kink at $\sim 70$~meV 
in the SC state~\cite{Kaminski-KinkNode-Bi2212,Kordyuk-KinkNodeEvolution-Bi2212}.
Beyond that energy, $1/\tau$ 
increases linearly with energy
in optimallly doped cuprates~\cite{Kaminski-AnisotropicSR}. 
A pressing issue in the
understanding of the physics of cuprates is how the scattering rate behaves
along the zone-edge (ZE), 
where the SC gap is maximum. An experimental difficulty
in h-doped cuprates is that the band along the ZE is very shallow
(about 50-100~meV), 
and the effects of interactions cannot be followed over a large energy range.
Thus, only a few experimental reports exist, and the debate is not
settled~\cite{Gromko-AntiNodalSCKink,Cuk-AntiNodalKink-B1g}.
In contrast, in e-doped cuprates, 
the ZE band-width is $\sim 500$~meV,
offering complete access to the momentum dependence 
of the many-body interactions. 
However, this kind of study has been scarcely addressed~\cite{Armitage-SelfEnNCCO}.

%
In this Letter we present a $k$-dependent study of the scattering rate in 
optimally doped Sm$_{1.85}$Ce$_{0.15}$CuO$_{4}$ (SCCO).
Compared to other e-doped cuprates, SCCO has the advantage of being cleaveable,
thus yielding a surface that is adequate for accurate ARPES studies.
The main results are as follows. First, we find kinks in both
the quasi-particle dispersion and scattering rate at 150~meV along the zone diagonal
and 70~meV along the zone edge.  Second, the scattering rate
along the diagonal increases with an approximately linear 
$\omega$-dependence beyond 150~meV.
In contrast, along the zone-edge 
the scattering rate saturates for $\omega > 200$~meV.
These results suggest that the electron interactions are highly anisotropic, 
being strong down to the bottom of the band along the diagonal, 
but becoming very weak beyond 200~meV along the zone-edge.

%
High-quality single crystalline SCCO samples were grown by a flux method and 
then annealed under low-oxygen pressure to render them SC with 
a $T_c=19$~K~\cite{Greene-XtalGrowth}.
Wavelength dispersive X-ray analysis gave a Ce concentration
$x=0.15 \pm 0.01$.
The ARPES experiments were 
done at the Synchrotron Radiation Center (SRC, University of Wisconsin, Madison) and 
the Swiss Light Source (SLS, Paul-Scherrer Institut, Switzerland)
using 55~eV linear and circular photons respectively.
A Scienta-2002 detector was used in both cases, with an
angular resolution of $0.25^{\circ}$. The energy resolutions
were 30~meV at SRC and 20~meV at SLS.
The samples were cleaved {\it in-situ} at 11~K in pressure better than 
$6\times10^{-11}$~Torr, and kept at these conditions during the measurements.
An eventual SC gap of $\sim 1-2$~meV~\cite{Matsui-SCgap} 
could not be resolved. The results were reproduced in three samples.
%
\begin{figure}
  \begin{center}
    \includegraphics[width=8cm]{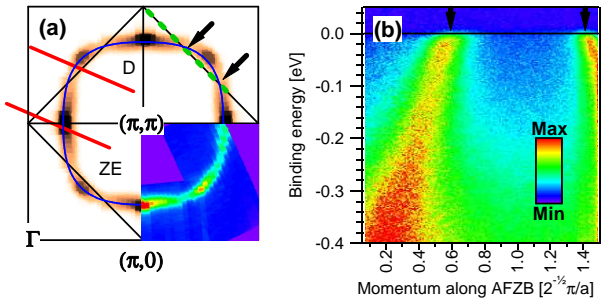}
  \end{center}
  \caption{\label{Fig1}
         (Color online) 
         (a) FS map of SCCO samples:
         Data for determining the FS was taken over the first 
         and second Brillouin zones.
         The spectra were integrated within 40~meV around $E_{F}$.
		 The lower-right quadrant displays a subset of such a raw data.
         The rest of the figure was obtained by symmetry operations, averaging
		 for equivalent points across the BZ diagonals, and saturating 
		 to white for intensities below 7\% of the maximum intensity
		 along the FS.
		 The red lines indicate cuts along the ZE and D directions discussed in the text. 
         The blue solid line is a single-band tight-binding fit to 
         the FS.
         (b) Dispersion along the AFZB [green dashed line in~(a)], 
         showing the absence of antiferromagnetic-induced folding. 
         The arrows indicate FS crossings. 
         }
\end{figure}

%
%
We first characterize the band structure of the studied samples.
Figure~\ref{Fig1}(a) shows the measured Fermi surface (FS),
chosen to be centered at $(\pi,\pi)$.
The data show a reduction of spectral weight around 
the crossings of the FS with the antiferromagnetic zone boundary (AFZB). 
The dispersion along the AFZB, 
shown in Fig.~\ref{Fig1}(b), shows only the bands corresponding to the 
barrel-like FS centered at $(\pi,\pi)$,
indicating that no antiferromagnetic-induced folding 
is present~\cite{Richard-AG-and-OR-PCCO,Armitage-FSvsDoping-NCCO}.
Such a folding has been observed only
in as-grown non SC samples~\cite{Richard-AG-and-OR-PCCO}, 
or in underdoped reduced SC samples~\cite{Matsui-BandFolding-UndNCCO,Park-UndSCCO}.
We simultaneously fitted the FS and the high-energy part of the band-structure
using a single-band tight-binding model:
\begin{eqnarray} 
	E_{tb} & = & \mu + t_{1}(\cos k_{x} + \cos k_{y}) 
	+ t_{2}(\cos k_{x}\cos k_{y}) \nonumber \\ 
	& & + t_{3}(\cos 2k_{x} + \cos 2k_{y}), 
	\label{EqTB}
\end{eqnarray}
with $(\mu,t_{1},t_{2},t_{3}) = (-10,-590,337,-96.6)$~meV.
Such a model is a good approximation of the
band-theory result~\cite{Andersen-LDA-cuprates}, 
and our fitting parameters are close to the canonical ones~\cite{Optics-PCCO-Millis-Zimmers}.
The resulting FS, shown by the continuous blue line in Fig.~\ref{Fig1}(a),
encloses an area corresponding to an electron-doping of $x=0.15 \pm 0.01$, 
in agreement with the bulk nominal doping.
%
\begin{figure}
  \begin{center}
    	\includegraphics[width=8cm]{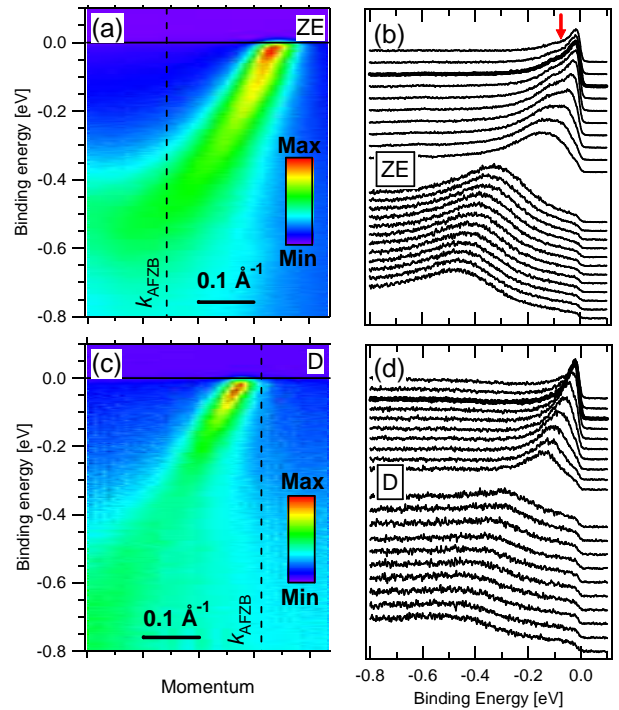}
  \end{center}
  \caption{\label{Fig2}
  		 (Color online)
         Data along the ZE~(a,b) and D~(c,d) directions of Fig.~\ref{Fig1}(a). 
         The intensity maps are shown as false-color plots in (a) and (c),
         with the AFZB wave vectors indicated by the vertical dashed lines. 
         EDCs near $k_{F}$ and near the bottom of the band 
         are shown in (b) and (d), with the EDCs at $k_{F}$ in bold.
         Along the ZE, the EDCs near $k_{F}$ show a peak-dip-hump structure, 
         with the dip located at 70~meV (arrow). The line-shape of the 
         high-energy EDCs along the ZE is approximately Lorentzian~$+$~background
         (see also Fig.~\ref{Fig3}).  In contrast, the high-energy EDCs along the
         diagonal are broad and asymmetric.
         }
\end{figure}
%

We now turn to the study of the many-body interactions. 
Within the sudden approximation, ARPES measures the occupied part
of the energy and momentum ($k$) dependent
single-particle spectral function $A(k,\omega)$~\cite{Hufner-PESbook}:
\begin{equation}
	A(k,\omega)=\frac{1}{\pi}\frac{\Sigma^{\prime\prime}(k,\omega)}
	{\left[\omega-\epsilon_{k}-\Sigma^{\prime}(k,\omega)\right]^{2}
	+\left[\Sigma^{\prime\prime}(k,\omega)\right]^{2}}.
\label{EqSpectralFn}
\end{equation}
Here, $\epsilon_{k}$ is the bare band structure, 
and $\Sigma = \Sigma^{\prime} + i \Sigma^{\prime\prime}$
is the complex self-energy (which quantifies the interactions).
Energy distribution curves (EDCs) are obtained when the photoemission intensity is 
plotted for constant $k$.  Momentum distribution curves (MDCs)
are obtained when it is plotted for constant $\omega$. 
One can obtain information about the self-energy
by analyzing the EDCs and MDCs~\cite{Damascelli-Review, Hufner-PESbook}. 
Thus, if $\Sigma$ is nearly $\omega$-independent for some extended
energy range, then the EDCs in this energy range become Lorentzians,
and their full-width at half-maximum (fwhm) is equal to $2\Sigma^{\prime\prime}$.  
Likewise, an MDC is a Lorentzian if, and only if, $\Sigma$ is independent of $k$ 
along the direction of the MDC and if the bare dispersion 
can be linearized in $k$. It then follows from Eq.~(\ref{EqSpectralFn})
that the position of the MDC peak, plotted as a function of $\omega$, 
gives the particle dispersion renormalized by interactions
$\epsilon^{\star}=\epsilon_{k}+\Sigma^{\prime}$,
and the MDC-fwhm is 
$\Delta k = 2\Sigma^{\prime\prime}/v_{k}$, where 
$v_{k}=d\epsilon_{k}/dk$ is the bare-band velocity --which is experimentally unknown.
Here, it is more appropriate to use the scattering rate $1/\tau$, defined 
in terms of measurable quantities as 
$1/\tau=\Delta k \times (d\epsilon^{\star}/dk)$~\cite{QPScatteringRate-Bi2212-Fujimori-Shen}.
%
\begin{figure}
  \begin{center}
    	\includegraphics[width=8cm]{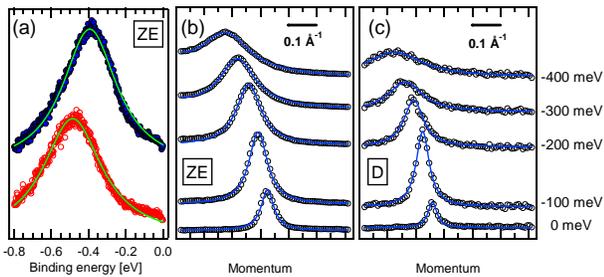}
  \end{center}
  \caption{\label{Fig3}
  		 (Color online)
         (a) Two selected EDCs along the ZE 
         (after linear background substraction),
         centered at $\omega \approx -400$~meV 
         (filled circles) and $\omega \approx -500$~meV (open circles), 
         and Lorentzian fits (solid lines).
         (b,c) Raw experimental MDCs (symbols) at selected binding energies along 
         the ZE and D cuts, respectively, and Lorentzian fits (lines).
         Along the ZE, the MDCs at $\omega \leq -300$~meV were fitted with
         two Lorenztians of equal width, to account for the peak of the
         left-branch of the dispersion (outside the experimental momentum window,
         but close enough to the right branch to create the observed assymetry).
         }
\end{figure}
%

\begin{figure}
  \begin{center}
     \includegraphics[width=8cm]{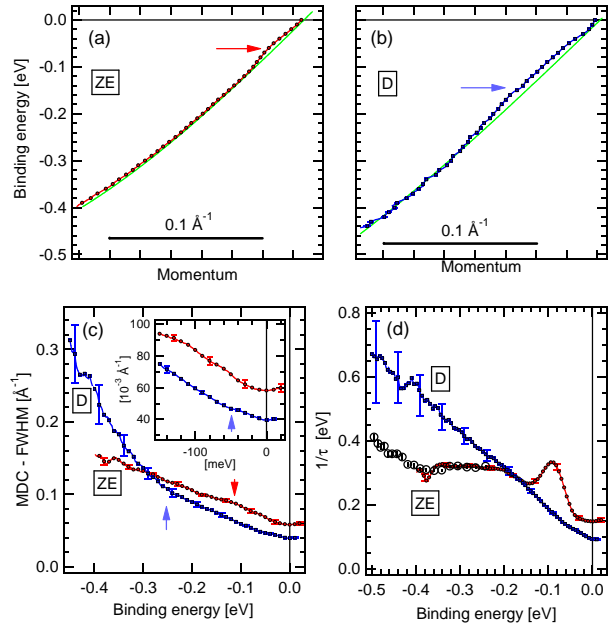}
  \end{center}
  \caption{\label{Fig4}
  		 (Color online)
         (a,b) MDC dispersions (symbols~$+$~lines) and tight-binding fits from Eq.~\ref{EqTB}
         (green lines) along the ZE and D cuts, respectively.  The
         scales are the same for comparison. 
         (c) MDC-fwhm as a function of binding energy
         for the ZE and D dispersions.
         The inset is a zoom over the low-energy part, showing a kink
         at about 50~meV along D. 
         (d) Scattering rate of the carriers (see text) along the ZE (circles) 
         and D (squares) cuts.
         The open circles represent the EDC-fwhm from 
         Lorentzian fits to the ZE data at high binding energies [Fig.~\ref{Fig3}(b)].
         In all plots, the error bars at selected points represent $\pm \sigma$ (the
         standard deviation) from the Lorentzian fit to the peak position 
         [Figs.~(a) and (b)] and fwhm [Figs.~(c) and (d)].
         When not apparent, the error bars are smaller than the symbol size.
         }
\end{figure}

Figures~\ref{Fig2}-\ref{Fig4} show the high-statistics data and analysis
along the ZE and D cuts indicated in Fig.~\ref{Fig1}(a).
Along the ZE direction [Figs.~\ref{Fig2}(a,b)], the dispersion 
of the EDC peak can be followed down the bottom
of the band, located at about 500~meV.
As seen in Fig.~\ref{Fig2}(b) 
the EDCs at $k_{F}$ along the ZE direction show a peak-dip-hump structure,
with the dip located at $\sim 70$~meV. 
For $\omega \gtrsim 200$~meV,
we find that the EDCs along this direction are well described by Lorentzians 
(with an approximately linear background) 
as shown in Fig.~\ref{Fig3}(a). 
All these features show that, along the ZE,
the band renormalization due to interactions is  
strong near $E_{F}$, but becomes very weak 
for $\omega \gtrsim 200$~meV,
pointing to an $\omega$-independent scattering rate, as
will be confirmed further.
By contrast, along the D direction, the EDCs around $k_{F}$
show a single peak that broadens rapidly upon dispersing 
towards the bottom of the band [Fig.~\ref{Fig2}(d)].
Thus, at low energies, the coupling strength to other excitations
appears to be weaker along the D than along the ZE.
However, for $\omega \gtrsim 100$~meV [Fig.~\ref{Fig2}(d)], 
the EDCs along the D direction become 
broad and asymmetric, 
suggesting that interactions 
along this direction 
become strong and remain $\omega$-dependent down to the bottom of the band.
On the other hand, 
the MDCs along both directions [Figs.~\ref{Fig3}(b,c)] are Lorentzians
down to $\omega \sim 400$~meV,
validating the MDC analysis that follows.
%

Figures~\ref{Fig4}(a) and (b) show the dispersions 
from the peak positions of the MDCs (symbols) 
along the ZE and D cuts, respectively.
The tight-binding band from Eq.~\ref{EqTB} for each direction is also shown
(green lines).
The arrows mark the energies where the experimental dispersions
are farthest from the tight-binding band, defined as the kink positions.
The raw MDC widths are shown in Fig.~\ref{Fig4}(c).
Along the ZE, the dispersion presents a kink at $\sim 60-70$~meV,
corresponding to the dip in the EDCs near $k_{F}$ [Fig.~\ref{Fig2}(b)].
A drop in the raw ZE linewidths [Fig.~\ref{Fig4}(c), full red circles] 
below a slightly larger energy of $\sim 100$~meV
is also observed, 
and is marked by an arrow~\cite{CommentKK}. 
Along the D cut, the experimental dispersion deviates 
from the tight binding fit at about 30~meV, then both dispersions run
almost parallel in the range $\sim 50-200$~meV, and merge
at $\sim 250$~meV. This broad kink structure, 
centered at $\sim 150$~meV [Fig.~\ref{Fig4}(b)],
is a feature not reported before in e-doped cuprates.
The raw D linewidths [Fig.~\ref{Fig4}(c), full blue squares]
present a slight kink at about 50~meV (shown in the inset). 
At larger energies, the D linewidths increase 
and present an up-turn at $\sim 250$~meV, about the same energy
at which the D dispersion and the tight-binding band rejoin.

The scattering rates along the ZE and D directions are shown in Fig.~\ref{Fig4}(d).
Along the ZE direction, $1/\tau$ shows a peak at about 80~meV 
(related to the kink and drop seen in the dispersions and linewidths, respectively),
while at energies larger than about 
200~meV $1/\tau$ practically saturates, indicating that 
the electron-scattering mechanisms become weakly or not energy-dependent. 
An energy-independent self-energy implies Lorentzian EDCs, 
as we actually observed [Figs.~\ref{Fig2}(b) and \ref{Fig3}(a)].
Moreover, as expected from an energy-independent self-energy, 
the EDC widths along the ZE in this energy range, shown by the 
open circles in Fig.~\ref{Fig4}(d), match very well with
$1/\tau$, which should then be close to the actual value of
$2\Sigma^{\prime\prime}$.
Along the D direction, on the other hand, a change of slope in
$1/\tau$ is seen at about 150~meV 
[Fig.~\ref{Fig4}(d)], while at larger 
energies this quantity continues to grow in an approximate linear way. 
This indicates that the scattering mechanisms
along the D direction are 
energy-dependent down to at least 500~meV, in agreement with our previous discussion on 
the EDCs along the D cut.  

We now compare our data with other ARPES data for e- and h-doped cuprates.
The structures at 70~meV along the ZE and 50~meV along the D
reproduce previous results on optimally doped SCCO, Nd$_{1.85}$Ce$_{0.15}$CuO$_{4}$ 
(NCCO), and Eu$_{1.85}$Ce$_{0.15}$CuO$_{4}$, 
and have been recently ascribed to electron-phonon 
interaction~\cite{Armitage-SelfEnNCCO,Park-SelfEnSCCO}.
At higher energies, our data along the D are in qualitative agreement 
with earlier data on NCCO~\cite{Armitage-SelfEnNCCO} 
that found that $\Sigma^{\prime\prime} \sim \omega^{\alpha}$, 
with $\alpha=1$ to 1.55, in the range $90<\omega<400$~meV.
Our results complete the picture for e-doped cuprates, showing that along the ZE
the interactions beyond 200~meV become weak, and the scattering rate
energy-independent.
As for the h-doped cuprates, a striking qualitative similarity with our data is the 
approximately linear form of the scattering rate 
along the diagonal~\cite{Damascelli-Review,Kordyuk-KinkNodeEvolution-Bi2212}.
However, we do observe the bottom of the D band,
so that no high-energy anomaly of the type reported in
h-doped cuprates~\cite{Lanzara-HEK, Johan-HEK} exists in e-doped cuprates.
Note also that, in h-doped cuprates, a kink at an energy of 150~meV
was recently unveiled, its physical origin 
being not yet clear~\cite{Zhou-Kink150meVBi2212}. Thus, our data suggest
the existence of what could be a new form of electron coupling that
is common to h- and e-doped cuprates.

Other probes in e-doped cuprates 
show signatures of the energy scales discussed here:
infrared data on 
Pr$_{2-x}$Ce$_{x}$CuO$_{4}$ at $x=0.13$ ($0.15$) show a broad  drop at 
$\sim 200$~meV (100~meV) in the optical scattering rates \cite{Zimmers-OpticsPCCO}.
Local-tunneling spectroscopy data on optimally-doped
SCCO show a hump at bias voltages $\sim 60-80$~meV \cite{Zimmers-TunnelSCCO}.
Both of these techniques are integrated over the FS.  
The optical response in cuprates is believed to be most sensitive to the excitations 
close to the diagonal, while the tunneling current is believed to be
most sensitive to the $(\pi,0)$ regions.
Thus, the infrared and tunnel results could be
reinterpreted in terms of the energy scales found in our angle-resolved data. 

In conclusion, among the $k$-space anisotropies in the carrier-dressing
shown by our data, our most significant observation is that
the scattering rate along the ZE 
saturates for binding energies larger than  $\sim 200$~meV, while
along the D direction $1/\tau$ increases with the binding energy $\omega$,
having a nearly linear $\omega$-dependence for $\omega \sim 150-500$~meV.
These results point to the interaction of electrons with a strongly dispersive
excitation with both a characteristic energy and a coupling strength to carriers
that are momentum dependent, such that interactions along the diagonal are strong
down to the bottom of the band, while they become very weak beyond $\sim 200$~meV
along the ZE.
In the framework of cuprate superconductivity,
two possibilities arise: either this unique scattering mechanism along the ZE
is also present in the h-doped cuprates, or the strong antiferromagnetic
fluctuations in e-doped cuprates separate two different regions
in $k$-space, with a physics describing the D region that is similar in h-
and e-doped cuprates, and a different picture describing the ZE region.

We acknowledge the European Union for supporting our work 
at the Swiss Light Source. The Synchrotron Radiation Center, 
University of Wisconsin-Madison, is supported by the National Science Foundation 
under award no. DMR-0537588. The work at the University of Maryland 
is supported by NSF contract DMR 0653535.
Work at Ames Laboratory was supported by the
Department of Energy - Basic Energy Sciences under Contract
No. DE-AC02-07CH11358.  
AFSS thanks LPEM-CNRS for financial support.
%

\end{document}